\documentclass[twocolumn,prb,superscriptaddress,showpacs]{revtex4}
\usepackage{graphicx}
\usepackage{dcolumn}

\begin{document}

\title{Ab initio study of origin and properties of a metal-organic interface
state of the PTCDA/Ag(111) system}

\author{N.~L.~Zaitsev}

\email{nza@yandex.ru}

\affiliation{Department of Physics, Ufa State Aviation Technical University, 450000 Ufa, Russia}

\author{I.~A.~Nechaev}

\affiliation{Department of Theoretical Physics, Nekrasov Kostroma State University, 156961 Kostroma, Russia.}

\affiliation{Research-Education Center ``Physics and Chemistry of High-Energy Systems'', Tomsk State
University, 634050 Tomsk, Russia}

\author{P.~M.~Echenique}

\author{E.~V.~Chulkov}

\affiliation{Donostia International Physics Center (DIPC), P. de Manuel Lardizabal,
4, 20018, San Sebastián, Basque Country, Spain}

\affiliation{Departamento de Física de Materiales, Facultad de Ciencias Químicas,
UPV/EHU and Centro Mixto CSIC-UPV/EHU, Apdo. 1072,\\
 20080 San Sebastián, Basque Country, Spain}

\begin{abstract}
We present a detailed study of a monolayer film of 3,4,9,10-perylene-tetracarboxylic acid dianhydride (PTCDA)
on Ag(111) (the PTCDA/Ag(111) system). The study is done within density functional theory with the use of the
periodic slab model. The slab is chosen to contain a PTCDA monolayer film on a silver thin film of different
thicknesses (6, 9, and 12 layers) with the (111) orientation. We show that one of two surface states of the
pure Ag(111) films transforms into an unoccupied interface state due to the adsorbate-substrate interaction.
The relation of the resulting state to the unoccupied state that has been experimentally observed in the
PTCDA/Ag(111) system by scanning tunneling and two photon photoemission spectroscopy is discussed.
\end{abstract}
\maketitle

\section{Introduction}

Theoretical and experimental investigations of electronic properties of organic molecular structures adsorbed
on metal surfaces are of great importance for the solution of both fundamental and practical problems
\cite{kahn_electronic_2003, barlow_complex_2003}. Among organic molecules, the particular research interest
is expressed in 3,4,9,10-perylene-tetracarboxylic acid dianhydride (PTCDA) molecules, since the latters are
able to form strictly ordered layers on the surfaces of some
metals.\cite{glckler_highly_1998,tautz_structure_2007} When adsorbed on Ag(111), the PTCDA monolayer film
causes the appearance of a delocalized two-dimensional band state that in
Ref.~\onlinecite{temirov_free-electron-like_2006} has been revealed at $\sim 0.7$ eV above the Fermi level
within scanning tunneling spectroscopy (STS) observations. This unoccupied state is described by a parabolic
dependence on the two-dimensional electron wave vector $\mathbf{k}$ with the effective mass
$m^{\ast}=0.47m_{e}$, where $m_{e}$ is the free electron mass. Aiming at understanding the mechanism of
electron delocalization in molecular structures, in spite of the similarity of $m^{\ast}$ with that of the
Shockley surface state (SS) of the pure Ag(111) surface, the authors of the cited work has attributed the
origin of such a dispersing state to the substrate-mediated interaction of LUMO+1 orbitals of ordered arrays
of the PTCDA molecules. The main experimental evidence of it is that the spatial distribution of the state is
concentrated on the molecules and resembles that of the LUMO+1 level of PTCDA. At that the anhydride groups
of PTCDA are involved in the wave function of the state.\cite{temirov_free-electron-like_2006}

In Ref.~\onlinecite{schwalb_electron_2008}, the unoccupied state of the PTCDA/Ag(111) system has been
investigated by means of two-photon photoemission spectroscopy (2PPES). The authors of the cited work have
presented experimental evidence that the state in question (here with the energy of 0.6 eV above the Fermi
level at $\mathbf{k}$ = 0 and with a parabolic dispersion with $m^{\ast}=0.39\pm0.03m_{e}$) comes from the SS
of Ag(111). The argumentation in favor of such an origin of the state is that their angle-resolved 2PPES
observations reveal the effective electron mass identical to the SS of the pure Ag(111) surface and the
location of the state at the metal-organic interface with large bulk penetration comparable to that of the
Shockley state. Thus, there are at least two diametrically opposite points of view on the origin of the
dispersing unoccupied state in the PTCDA/Ag(111) system. Since such states play a crucial role in the
dynamics of excited electrons and charge transfer through the interface \cite{chulkov_excitation_2006,
wusten_hot-electron_2008}, the question of the origin and properties of the unoccupied state in the
PTCDA/Ag(111) system is important to be cleared up.

In this paper, within the periodic slab model we study the electronic structure of the PTCDA/Ag(111) system,
using \emph{ab initio} density functional theory calculations. The slab is chosen to contain the PTCDA
monolayer film on Ag(111) films of different thickness (6, 9, and 12 layers). In the monolayer, the PTCDA
molecules are packed in a herringbone structure. On the one hand, we show that the experimentally observed
unoccupied state arises from the Shockley surface state, has the bulk penetration similar to that of the
latter state, and is localized at the interface. On the other hand, we demonstrate that in the plane of the
PTCDA monolayer film the spatial distribution of this interface state is concentrated on the molecules with
quite big amplitude on the perylene backbone and anhydride groups and relatively small -- at the perylene
edges of the PTCDA molecule. Thus, we corroborate the interpretation proposed in
Ref.~\onlinecite{schwalb_electron_2008} and explain the STS observations of
Ref.~\onlinecite{temirov_free-electron-like_2006}.

\section{Calculation details}

We use the periodic slab model to study the metal-organic interface. The electronic structure calculations
were performed by using the OPENMX code,\cite{ozaki_numerical_2004,ozaki_efficient_2005} which are based on
the density functional theory (DFT) and a linear combination of localized pseudoatomic orbitals method. We
exploit Troullier-Martins type pseudopotentials\cite{troullier_efficient_1991} with a partial core correction
to replace the deep core potentials by norm-conserving soft potentials. Upper core electrons for the Ag
pseudopotential are included in addition to valence electrons in order to take into account the contribution
of semicore states to the electronic structure. As a basis set, we take double-valence plus
single-polarization orbitals. For hydrogen, carbon and oxygen atoms constituting PTCDA, variationally
optimized basis orbitals for biological molecules are used with cut-off radii of 5.0 for C and 4.5 a.u. for O
and H atoms.\cite{ozaki_variationally_2004} For Ag atoms cut-off radius is 7.5 a.u. The real-space grid
techniques are used with the energy cutoff of 190 Ry in numerical integrations and solution of the Poisson
equation with the help of the fast Fourier transformation. The total-energy convergence was better than 0.027
meV.

\begin{figure}[tbp]
\includegraphics[angle=-90, scale=0.35]{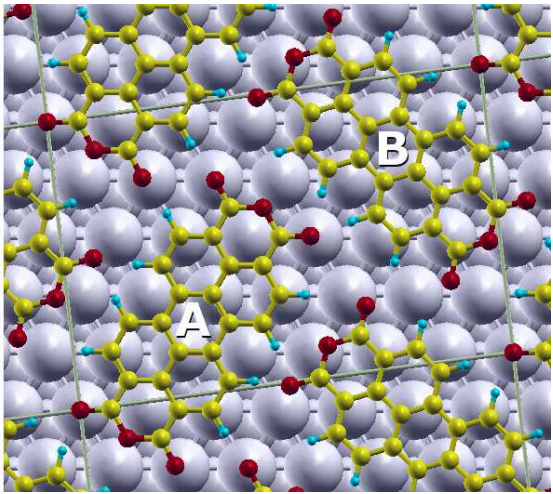}
\caption{\label{fig:unit_cell} (Color online) Structure of the PTCDA monolayer on the Ag(111) surface.}
\end{figure}

To describe the PTCDA/Ag(111) system, we employ a slab containing a silver film with the (111) orientation
and $N_{l}$ atomic layers ($N_{l}$=6, 9, 12) together with a PTCDA monolayer film, which is ``applied'' on
one side of the silver film only, and a vacuum region corresponding to 6 silver interlayer spacings. The
distance between the silver surface layer and the PTCDA monolayer is taken to be equal to experimental value
of 2.86 \AA{} (see Refs.~\onlinecite{hauschild_molecular_2005} and \onlinecite{henze_vertical_2007}). A close
value can be obtained in DFT calculations\cite{rohlfing_adsorption_2007} with exchange-correlation potential
in the local density approximation (LDA), while in the generalized gradient approximation the distance
becomes very large ($\sim3.5$ \AA{} in Ref.~\onlinecite{hauschild_molecular_2005} and
\onlinecite{rurali_commentmolecular_2005}) and the adsorption energy is underestimated. This fact causes the
use of the LDA in our DFT calculations.

The PTCDA monolayer on Ag(111) has an ordered periodic structure that consists of two types (denoted as A and
B in Fig.~\ref{fig:unit_cell}) of molecules misoriented by an angle of $77^{\circ}$ with respect to each
other (a herringbone pattern). The A-type molecule is aligned parallel to the row of silver atoms, while the
B-type molecule is misaligned by an angle of $17^{\circ}$ (see Refs.~\onlinecite{kraft_lateral_2006} and
\onlinecite{rohlfing_adsorption_2007}). The superstructure of the interface is given by a nearly rectangle (a
deviation of $1^{\circ}$) unit cell of $18.96\times12.61$ \AA{}$^{2}$. The resulting periodic supercell of
the slab model contains $33\times N_{l}$ silver atoms (33 -- the number of atoms per layer) and 76 atoms
belonging to two PTCDA molecules. The Brillouin zone of the supercell is sampled with a ($3\times3\times1$)
mesh of $\mathbf{k}$ points. The $z$-axis is directed normal to the surface of the films.

\section{Results and discussion}

We start with an band-structure analysis of clean Ag(111) films with the supercell that contains $33\times
N_{l}$ silver atoms in order to identify surface states (SS) in the absence of the energy gap at the
$\overline{\Gamma}$ point of the corresponding small surface Brillouin zone and to make sure that
characteristics of these states are the same as in the case of the supercell with $1\times N_l$ silver atoms,
which gives the familiar surface band structure for the Ag(111) surface. Note that in the case of a film
there are two surface states with different energies caused by bonding-antibonding splitting due to
interaction through the film. We distinguish these two states as SS1 and SS2 with energies $E_{SS1}>E_{SS2}$
at the $\bar{\Gamma}$ point. A comparison has shown that the surface-state energies obtained with the
$33\times N_{l}$ supercell (see Table \ref{tab:SS-IS}) coincide with those found with the $1\times N_{l}$
supercell (not presented) within $\sim10$ meV, while the bonding-antibonding splitting ($\Delta E_{clean}$)
--- within $\sim2$ meV. As regards the effective masses, their values practically do not vary with the number
of atoms per layer.

\begin{table}
\centering \caption{\label{tab:SS-IS} Energies $E_{SS1}$ and $E_{SS2}$ of the surface states at the
$\overline{\Gamma}$ point and corresponding effective masses $m^{\ast}_{SS1}$ and $m^{\ast}_{SS2}$ for clean
Ag(111) films with different number of layers ($N_l$). $\Delta E_{clean}=E_{SS1}-E_{SS2}$ and
$E_{clean}^{av}=E_{SS2}+\Delta E_{clean}/2$. In the case of the PTCDA monolayer on the Ag(111) films,
$E_{IS}$ and $E_{SS}$ are the $\overline{\Gamma}$-point energies of the interface and surface state with
effective masses $m^{\ast}_{IS}$ and $m^{\ast}_{SS}$, respectively. $\Delta E=E_{IS}-E_{SS}$ and
$E_{av}=E_{SS}+\Delta E/2$. All the energies are measured electron volts from the Fermi energy.}

\begin{ruledtabular}
\begin{tabular}{ccccccc}
\multicolumn{7}{c}{Ag(111)} \tabularnewline \hline $N_{l}$  & $E_{SS1}$  & $m_{SS1}^{\ast}$  & $E_{SS2}$ &
$m_{SS2}^{\ast}$  & $\Delta E_{clean}$  & $E_{clean}^{av}$\tabularnewline

6  & 0.141  & 0.40  & -0.460  & 0.53  & 0.601  & -0.159\tabularnewline

9  & 0.022  & 0.38  & -0.227  & 0.46 & 0.250  & -0.102\tabularnewline

12  & -0.023  & 0.38  & -0.135  & 0.42 & 0.112  & -0.079\tabularnewline

\hline \multicolumn{7}{c}{PTCDA/Ag(111)}\tabularnewline \hline $N_{l}$  & $E_{IS}$  & $m_{IS}^{\ast}$  &
$E_{SS}$  & $m_{SS}^{\ast}$  & $\Delta E$ & $E_{av}$\tabularnewline \hline

6 & 0.523  & 0.43  & -0.271  & 0.76  & 0.694 & 0.176\tabularnewline

9  & 0.469  & 0.42  & -0.119  & 0.46  & 0.588  & 0.175\tabularnewline

12 & 0.474  & 0.46 & -0.087  & 0.42  & 0.561  & 0.193\tabularnewline
\end{tabular}
\end{ruledtabular}
\end{table}

The mentioned characteristics of the surface states depend on thickness of the film (see Table
\ref{tab:SS-IS}). As was expected, $\Delta E_{clean}$ decreases rather rapidly with increasing $N_l$ and
should vanish at the $N_l\rightarrow\infty$ limit. The energy $E_{SS1}$ tends to reduce, while $E_{SS2}$
seeks to increase. At the same time the effective mass $m_{SS1}^{\ast}$ varies weakly with the thickness as
compared to $m_{SS2}^{\ast}$. Note that as $N_l$ increases the average energy $E_{clean}^{av}$ becomes closer
to the surface-state energy $E_{SS}^{exp}=-63\pm1$ meV experimentally observed\cite{reinert_direct_2001} for
the Ag(111) surface and already for the twelve-layer film $E_{clean}^{av}$ differs from $E_{SS}^{exp}$ within
$\sim16$ meV.

 \begin{figure*}[tbp]
 \includegraphics[angle=0,scale=0.95]{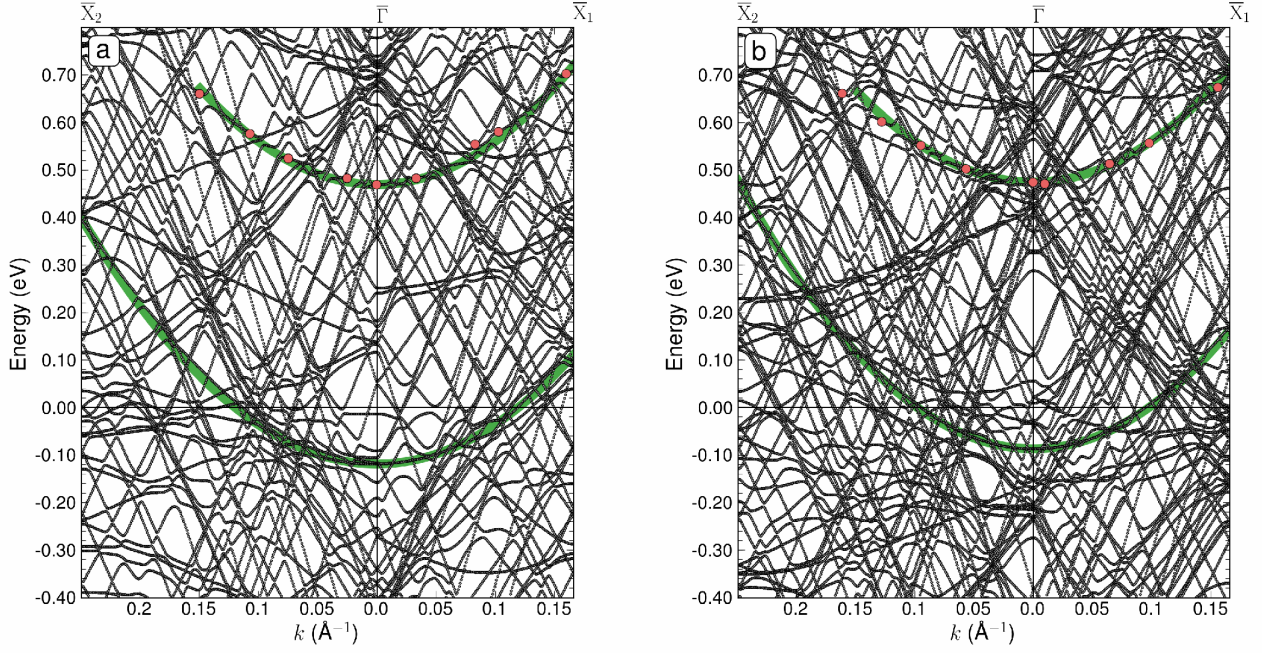}
 \caption{\label{fig:p9ml_bands}(Color online) Band structure of the PTCDA monolayer on the nine- (a) and
 twelve-layer (b) Ag(111) film in the $\bar{X}_{2}\rightarrow\bar{\Gamma}\rightarrow \bar{X}_{1}$ direction of
 the surface Brillouin zone. The thick gray lines show the parabolic approximation of the dispersion of the
 interface (upper line) and surface (lower line) states. Points indicate wave vectors $\mathbf{k}$, at which
 the spatial distribution of the interface-state wave function is analyzed (see Table \ref{tab:charges} and
 Fig. \ref{fig:z-chg}).}
 \end{figure*}

Now we consider the PTCDA monolayer on the Ag(111) film of different thicknesses. Fig.~\ref{fig:p9ml_bands}
shows the obtained band structure for two cases: nine- and twelve-layer silver films. Also, the figure
reflects our results on analysis of the square modulus of wave functions of found states (see the thick
lines). When averaged over spatial coordinates $xy$ at different $\mathbf{k}$ along
$\bar{\Gamma}\rightarrow\bar{X}_{2}$ and $\bar{\Gamma}\rightarrow\bar{X}_{1}$ directions, the square modulus
can be considered as a function of $z$, which gives important information about charge-density distribution
[denote as $\rho(z,\mathbf{k})$] along the direction normal to the surface. We have thus revealed the partly
occupied surface state (the lower thick line in Fig.~\ref{fig:p9ml_bands}) with the wave function localized
in the region of the ``clean'' surface of the Ag(111) film. At the $\bar{\Gamma}$ point, the energy (denote
as $E_{SS}$) of such a state depends on the film thickness and, as is seen from Table \ref{tab:SS-IS}, tends
to $E_{clean}^{av}$ with increasing $N_l$. This surface state is originated from the state marked as SS2 in
the case of the clean Ag(111) films. Note that in spite of the difference $E_{SS}-E_{SS2}$, which is caused
by the presence of the PTCDA monolayer and decreases with an increase of $N_l$ (see Table \ref{tab:SS-IS}),
the effective masses $m^{\ast}_{SS}$ and $m^{\ast}_{SS2}$ are the same (except for the six-layer silver
film). At that, for the twelve-layer silver film they become very close to the experimental value of
$0.40m_e$ for the surface-state effective mass found for the Ag(111) surface from photoemission data in
Ref.~\onlinecite{reinert_direct_2001}.

 \begin{figure*}[tbp]
 \includegraphics[angle=0,scale=0.95]{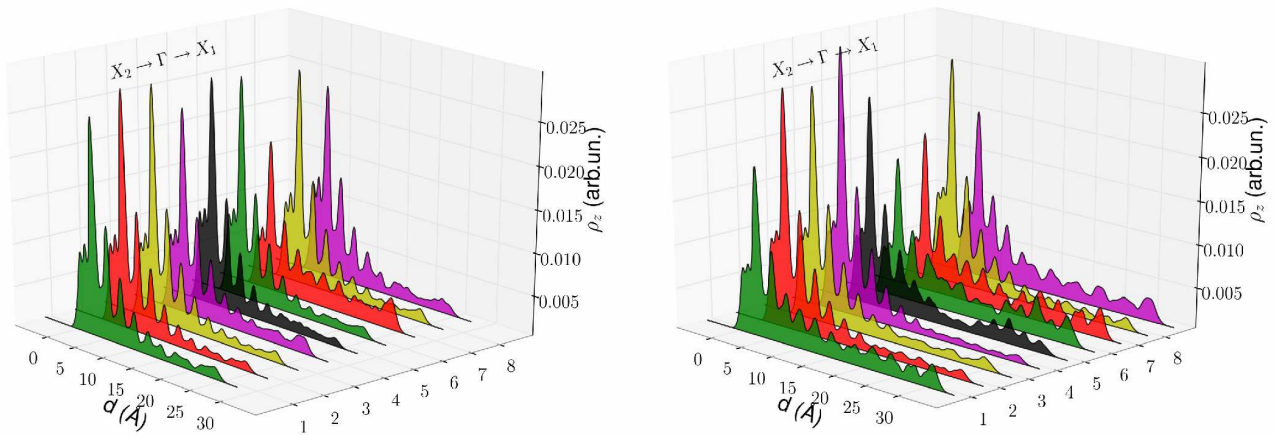}

 \caption{\label{fig:z-chg}(Color online) Unity-normalized charge-density distribution
 $\rho_{IS}(z,\mathbf{k})$ for the interface states as a function of $z$ for different $\mathbf{k}$ in the
 $\bar{X}_{2}\rightarrow\bar{\Gamma}\rightarrow\bar{X}_{1}$ direction. (a) The PTCDA monolayer on the
 nine-layer Ag(111) film. (b) The PTCDA monolayer on the twelve-layer Ag(111) film. The distributions for
 different $k$-points shown in Fig. \ref{fig:p9ml_bands} are presented in order from left to right in
 correspondence with $k$-points listed in Table \ref{tab:SS-IS} in order from top to bottom.}
 \end{figure*}

On analyzing $\rho(z,\mathbf{k})$, also we can unambiguously say that in the spectrum above the Fermi level
there is an interface state (the upper thick line in Fig. \ref{fig:p9ml_bands}) whose wave function has a
pronounced maximum in the region of the metal-organic interface (see Fig. \ref{fig:z-chg}) and a
surface-state-like penetration into the silver film. Actually, as is seen from Table \ref{tab:charges}), more
than 50\% of the IS charge accumulates in the metal film. The rest of the charge are divided among the
interface and molecular-monolayer regions as $\sim2/3$ and $\sim1/3$, respectively. As to the energy $E_{IS}$
of the interface state (IS) at the $\bar{\Gamma}$ point, we note that $E_{IS}$, which is about 0.5 eV, is
quite stable with respect to the film thickness (see Table \ref{tab:SS-IS}). The dispersion of the interface
state is adequately described by a quadratic dependence on the wave vector $\mathbf{k}$ with an effective
mass $m^{\ast}_{IS}$, which is close to that for the aforemention surface state for $N_l>6$.

To all appearances, the presented analysis corroborates the interpretation of Ref.
\onlinecite{schwalb_electron_2008}, in which the unoccupied dispersing state of the PTCDA/Ag(111) system is
considered as a genuine interface state arising from an upshift of the occupied Shockley surface state of the
clean Ag(111) surface. In fact, as our results show, there is no other unoccupied state, which we could
associate with the state discussed in Refs. \onlinecite{schwalb_electron_2008} and
\onlinecite{temirov_free-electron-like_2006}. In our case, the only interface state we have found can be
considered as a transformation of the surface state that is marked as SS1 in the case of the clean Ag(111)
films. The transformation is caused by the interaction of the Ag(111) film with the PTCDA monolayer.

\begin{table}
\centering \caption{\label{tab:charges}Charges of the interface state in regions of the silver film
($Q_{b}$), the interface ($Q_{i}$), and the PTCDA monolayer film ($Q_{m}$) at different values of
$\mathbf{k}$ along the  $\bar{X}_{2}\rightarrow\bar{\Gamma}\rightarrow \bar{X}_{1}$ direction of the surface
Brillouin zone (see points in Fig. \ref{fig:p9ml_bands}). Two cases of nine- and twelve-layer silver films
are presented. To calculate the charges, the unity-normalized charge-density distribution
$\rho_{IS}(z,\mathbf{k})$ shown in Fig. \ref{fig:z-chg} has been used.}
\begin{ruledtabular} \begin{tabular}{cccccccc}
\multicolumn{4}{c|}{$N_l=9$} & \multicolumn{4}{c}{$N_l=12$}\tabularnewline \hline $k$, \AA$^{-1}$  &
$Q_{b},e$ & $Q_{i},e$  & $Q_{m},e$  & $k$, \AA$^{-1}$  & $Q_{b},e$  & $Q_{i},e$  & $Q_{m},e$\tabularnewline
\hline 0.150  & 0.53  & 0.30  & 0.17  & 0.161  & 0.66  & 0.22  & 0.12\tabularnewline 0.107  & 0.50  & 0.33  &
0.17 & 0.128  & 0.53  & 0.30  & 0.17\tabularnewline 0.075  & 0.51  & 0.32  & 0.17  & 0.095  & 0.55  & 0.29  &
0.16\tabularnewline 0.025  & 0.55  & 0.29  & 0.16  & 0.057  & 0.51  & 0.33  & 0.16\tabularnewline \hline 0.00
& 0.49  & 0.32  & 0.19  & 0.00  & 0.60  & 0.27  & 0.13\tabularnewline \hline 0.033  & 0.51  & 0.32  & 0.17  &
0.010  & 0.72  & 0.18  & 0.09\tabularnewline 0.083  & 0.66  & 0.22  & 0.12 & 0.065  & 0.70  & 0.20  &
0.09\tabularnewline 0.103  & 0.52  & 0.30  & 0.18  & 0.098  & 0.53  & 0.29  & 0.17\tabularnewline 0.160  &
0.53  & 0.27  & 0.20  & 0.156  & 0.65  & 0.22  & 0.13\tabularnewline
\end{tabular}\end{ruledtabular}
\end{table}

 \begin{figure}[tbp]
 \includegraphics[angle=0,scale=0.85]{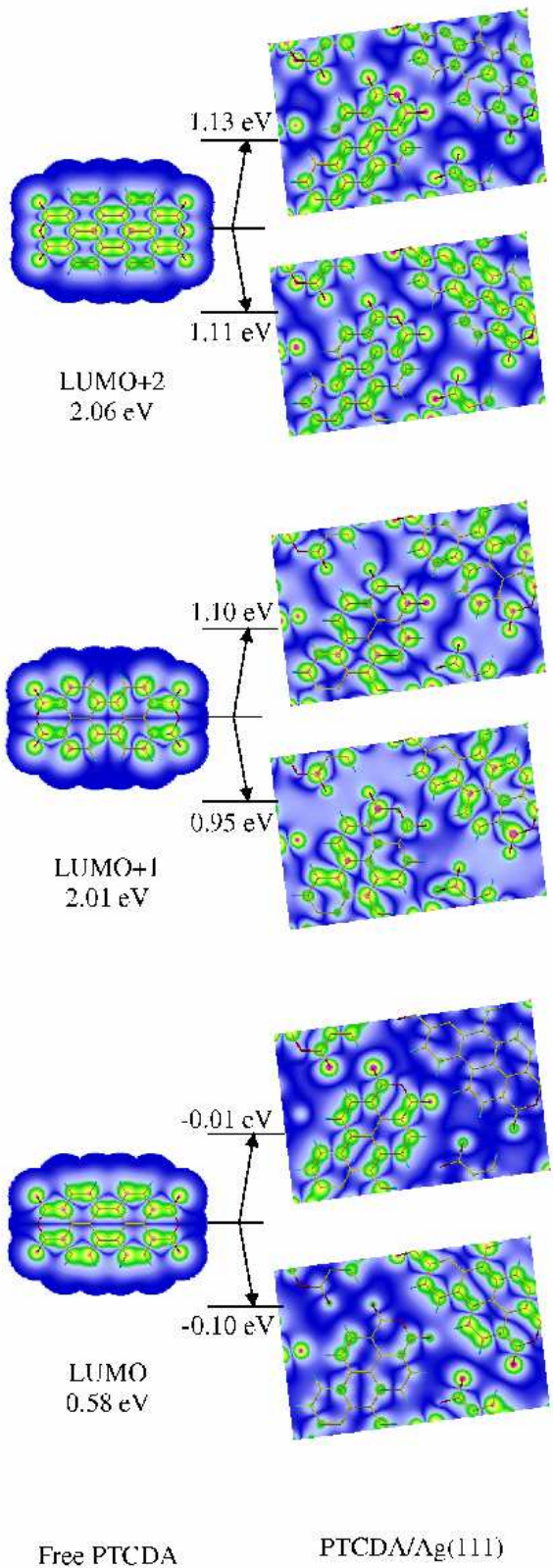}
 \caption{\label{fig:spatial_distribution}(Color online) Spatial distribution of three lowest unoccupied
 orbitals of the free PTCDA molecule (at the left) and corresponding states in the PTCDA/Ag(111) system at the
 $\bar{\Gamma}$ point in the case of the six-layer silver film (at the right).}
 \end{figure}

\begin{figure}[tbp]
 \includegraphics[angle=0,scale=0.4]{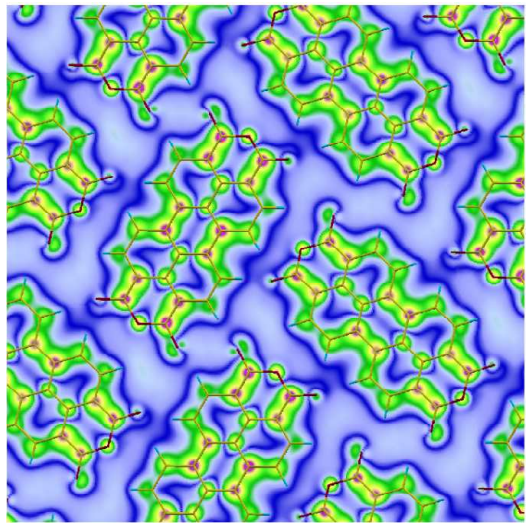}
 \caption{\label{fig:IS_sd} (Color online) Spatial distribution of the interface state in the PTCDA/Ag(111)
 system at the $\bar{\Gamma}$ point in the molecular plane for the case of the nine-layer silver film.}
 \end{figure}

Now, in order to establish a connection of our results with the STS observations of
Ref.~\onlinecite{temirov_free-electron-like_2006}, we should trace the transformation of molecular orbitals
upon moving from a free PTCDA molecule to the PTCDA/Ag(111) system. To this end, we analyze energies,
charge-density distributions $\rho(z,\mathbf{k}=0)$, and spatial distributions of corresponding states in the
molecular plane. We recognize electronic states in the PTCDA/Ag(111) system as those originated from
molecular levels from the localization of the wave function predominantly on the PTCDA monolayer.
Fig.~\ref{fig:spatial_distribution} shows our results of such an analysis. As follows from the figure, the
LUMO level of the free molecule [with the structure as it appears in the PTCDA monolayer on Ag(111)] goes
down under the Fermi level of the PTCDA/Ag(111) system and splits due to the presence of two types of
molecules (A an B) in the unit cell (see Fig.~\ref{fig:unit_cell}). At that the spatial distribution, which
is slightly modified under the influence of intermolecular interaction and adsorbate-substrate interaction,
preserves practically all its key features. Note that the wave function of the state with smaller energy is
predominantly localized on the B-type molecule, while the state with bigger energy is characterized by the
wave function, which is mainly localized on the A-type molecule. The nearly degenerate LUMO+1 and LUMO+2
orbitals demonstrate the same tendency. As well as in Refs. \onlinecite{rohlfing_adsorption_2007} and
\onlinecite{romaner_theoretical_2009}, the states originated from these orbitals fix their energies in the
vicinity of 1eV. An analysis has shown that the dispersion of these states are very small as compared with
that of the found interface state. As regards the spatial distribution of the latter, Fig. \ref{fig:IS_sd}
clearly shows that, as well as in the case of the states arising from the LUMO+1 and LUMO+2 orbitals, the IS
charge density is concentrated on the molecules. However, the character of the distribution on the molecule
is quite different. Moreover, in contrast to the states arising from the LUMO+1 and LUMO+2 orbitals, the IS
charge density is distributed more or less uniformly among the A- and B-type molecules.

The analysis presented above does not allow us to follow the interpretation of the unoccupied dispersing
state of the PTCDA/Ag(111) system, which was made in Ref. \onlinecite{temirov_free-electron-like_2006}, as
the state formed by the molecular orbitals. At that, such main features of the dispersing state, which have
been observed in the cited work by the STS, as the dominant localization on the perylene backbone and the
involvement of the anhydride groups in formation of the state are clearly seen in Fig. \ref{fig:IS_sd}. Here
we would like to note that in the molecular plane the spatial distribution of the IS does not vary with the
silver-film thickness and retains its form as the wave vector $\mathbf{k}$ changes along the symmetric
directions of the Brillouin zone.

\section{Conclusion}

In conclusion, we have investigated electronic structure of the PTCDA/Ag(111) system within density
functional theory with the use of the periodic slab model. The slab has been chosen to contain the PTCDA
monolayer on the thin Ag(111) film. We have considered the silver films of different thickness (6, 9, and 12
layers) to fix a trend in changes in the electronic structure of the Ag(111) film upon adsorption of the
PTCDA monolayer and to check the convergence of our results. We have found that upon the adsorption one of
two surface states of the silver film, the wave function of which is localized on the clean side of the film,
tends to have the $\bar{\Gamma}$-point energy as that for the clean Ag(111) surface. Another one transforms
into an unoccupied interface state with the wave function that has a pronounced maximum in the region of the
metal-organic interface. The interface state has the $\bar\Gamma$-point energy of $\sim0.5$ eV, which is
weakly varying (within $\sim50$ meV) with the silver-film thickness. The dispersion of this state is nearly
isotropic and adequately described by a quadratic dependence on the two-dimensional electron wave vector with
the effective mass $0.43m_{e}$, $0.42m_{e}$ and $0.46m_{e}$ for six-, nine-, and twelve-layer Ag(111) film,
respectively.

We have related the aforementioned interface state with the unoccupied dispersing state experimentally
observed in the PTCDA/Ag(111) system by scanning tunneling\cite{temirov_free-electron-like_2006} (the
effective mass of $0.47m_{e}$) and two photon photoemission\cite{schwalb_electron_2008} spectroscopy (the
effective mass of $0.39\pm0.03m_{e}$). We have demonstrated that as in the STS observations, in the plane of
the PTCDA monolayer film, the spatial distribution of the interface state is concentrated on the molecules
with quite large charge density on the perylene backboone and anhydride groups and relatively small -- at the
perylene edges of the PTCDA molecule. At that, as it has been revealed by the 2PPES, the wave function of
this state has a surface-state-like penetration into the silver film. We thus have corroborated the
interpretation of the origin of the unoccupied dispersing state of the PTCDA/Ag(111) system, which has been
proposed in Ref.~\onlinecite{schwalb_electron_2008}.

Finally, we would like to note that in our calculation the unoccupied interface state appears at lower energy
than it is experimentally observed (0.7 eV in Ref.~\onlinecite{temirov_free-electron-like_2006} and 0.6 eV in
Ref.~\onlinecite{schwalb_electron_2008}). It can be caused by using the LDA to describe electronic states of
the metal-organic interface.\cite{romaner_theoretical_2009} We believe that as in the case of the free PTCDA
molecule (see, e.g., Ref. \onlinecite{dori_valence_2006}) the $GW$ approximation can improve the description
of the electronic spectrum of the system under question.

\section*{\label{sec:acknowledgments}Acknowledgments}

We thank Tomsk State University and Ufa State Aviation Technical University for supercomputer time provided
for the computations.

\end{document}